\documentclass[conference]{IEEEtran}
\IEEEoverridecommandlockouts

\def\BibTeX{{\rm B\kern-.05em{\sc i\kern-.025em b}\kern-.08em
    T\kern-.1667em\lower.7ex\hbox{E}\kern-.125emX}}

\usepackage{cite}
\usepackage[T1]{fontenc}
\usepackage{graphicx}
\usepackage{booktabs}
\usepackage{amsmath,amssymb,amsfonts}
\usepackage[dvipsnames]{xcolor}
\usepackage{placeins}
\usepackage{dblfloatfix}
\usepackage{hyperref}
\newcommand{\HeI}{\mbox{He\,I 10830~\AA}}

\usepackage[compact]{titlesec}

\allowdisplaybreaks
\begin{document}
	\setlength{\abovedisplayskip}{5pt}
	\setlength{\belowdisplayskip}{5pt}
	\setlength{\abovedisplayshortskip}{2pt}
	\setlength{\belowdisplayshortskip}{2pt}

\title{Cross-Modal Solar Image Synthesis: Adapting the Surya Foundation Model from \HeI{} to EUV Translation and Coronal Hole Segmentation}

 
\author{\IEEEauthorblockN{Marco Marena\textsuperscript{1}, Andrés Muñoz Jaramillo\textsuperscript{2}, Qin Li\textsuperscript{1}, Haodi Jiang\textsuperscript{3}, Jinghao Cao\textsuperscript{1}, Wen He\textsuperscript{1}, \\ Ziyang Zhang\textsuperscript{1}, Chenxi Yuan\textsuperscript{1},  Chao Wang\textsuperscript{4}, Haimin Wang\textsuperscript{1}, Bo Shen\textsuperscript{1*} \thanks{\textsuperscript{*}Dr. Bo Shen is the corresponding author. Email: bo.shen@njit.edu} }
\IEEEauthorblockA{\textsuperscript{1}New Jersey Institute of Technology, Newark, NJ; \textsuperscript{2}SouthWest Research Institute, Boulder, CO} 
\IEEEauthorblockA{\textsuperscript{3}Sam Houston State University, Huntsville, TX; \textsuperscript{4}University of Iowa, Iowa City, IA} 
\vspace{-0.2in}}

\maketitle

\begin{abstract}
The long observational record of \HeI{} offers a means to investigate solar morphology before modern extreme-ultraviolet (EUV) imaging. We adapt the Surya solar foundation model to predict Solar Dynamics Observatory/Atmospheric Imaging Assembly (SDO/AIA) 94, 193, and 304~\AA{} images and a coronal hole (CH) probability map from full-disk helium observations. A convolutional input adapter, low-rank backbone updates, and dedicated output decoders learn from temporally paired, geometrically registered observations, with Spatial Possibilistic Clustering Algorithm (SPoCA) catalog polygons providing CH supervision. On observations held out from downstream fine-tuning, the selected dedicated models achieve disk-restricted correlation coefficients (CCs) of 0.8196, 0.8885, and 0.8398 for the three AIA channels, respectively; the CH model achieves an intersection over union (IoU) of 0.4360. The predictions recover broad solar structure, although local agreement varies substantially by channel. An optional residual refiner addresses spatial detail, and its application on pre-SDO dates improves the correlation of synthetic AIA 304 with Solar and Heliospheric Observatory/Extreme-ultraviolet Imaging Telescope (SOHO/EIT) 304 references from 0.6511 to 0.7085. Together, these results support the feasibility of helium-conditioned EUV morphological proxies and motivate their use in historical reconstruction, within the scope of the downstream test and cross-instrument evaluation.
\end{abstract}


\section{Introduction}\label{sec:intro}
 Reconstructing earlier solar activity requires connecting long ground-based records with the atmospheric diagnostics available from modern space missions. Direct extreme-ultraviolet (EUV) images reveal structures that are central to coronal studies, but their observational coverage is shorter than the record of several chromospheric proxies. A useful historical reconstruction would express information in those earlier observations in a consistent EUV representation, allowing broad bright and dark structures to be compared across eras. This motivates the question studied here: how much of the morphology in several EUV channels, and in coronal hole (CH) maps, can be predicted from full-disk \HeI{} observations?

The neutral-helium triplet provides a physically motivated input to this problem. The Synoptic Optical Long-term Investigations of the Sun (SOLIS) facility was designed for systematic observations, including He\,{\sc i} 1083.0~nm spectroheliograms \cite{Keller2003SOLIS,NSOSOLISDataProducts}. The triplet forms in the upper chromosphere, with its population influenced by photoionization from coronal and transition-region radiation followed by recombination into the triplet system; collisional excitation can also contribute \cite{AndrettaJones1997}. This coupling supports the use of He\,{\sc i} morphology as a proxy for large-scale coronal structure and for CH mapping \cite{HarveyRecely2002}. The mapping remains conditional because different atmospheric states can produce similar helium signatures.

The Solar Dynamics Observatory (SDO) provides sustained observations of solar variability, and its Atmospheric Imaging Assembly (AIA) images the full Sun at high cadence in seven EUV and three ultraviolet/visible passbands \cite{Pesnell2012,Lemen2012}. AIA passbands are broadband and generally multithermal, so they should not be interpreted as monochromatic temperature measurements. Their response functions nevertheless provide complementary diagnostics: 94~\AA{} includes a strong high-temperature response relevant to flares and hot active-region structures; 193~\AA{} has major coronal and high-temperature flare contributions; and 304~\AA{} is normally dominated by He\,{\sc ii} emission from the chromosphere and transition region \cite{ODwyer2010}. The three targets therefore provide a stringent test of whether He\,{\sc i} observations contain useful information about distinct coronal and transition-region morphologies.

The AIA 193~\AA{} channel is also widely used to identify a CH, an extended region of reduced EUV emission. CHs have low coronal density, are associated with predominantly open magnetic fields, and are important sources of high-speed solar-wind streams \cite{Cranmer2009,KristaGallagher2009}. Intensity alone does not define CH boundaries uniquely: filament channels, limb effects, and instrumental artifacts can produce similar dark signatures. We therefore use the Spatial Possibilistic Clustering Algorithm (SPoCA), an EUV-image segmentation and tracking suite that combines fuzzy/possibilistic clustering with spatial and temporal region processing \cite{Verbeeck2014SPoCA}, to provide the algorithmic CH labels. CH segmentation is consequently formulated as a probabilistic task distinct from EUV-intensity regression.

Image translation provides a way to learn this relationship during periods of overlapping observations \cite{asensio2023machine,shen2024deep}. Park et al. generated AIA ultraviolet (UV)/EUV images from Helioseismic and Magnetic Imager (HMI) magnetograms \cite{Park2019}; Son et al. generated He\,{\sc i} from AIA \cite{Son2021}. Other studies used H$\alpha$, He\,{\sc i}, and EUV products \cite{Liu2021,Marena2025,DiazBaso2018,Jarolim2025}, building on conditional translation methods \cite{Isola2017}. Our earlier diffusion-based He I-to-AIA 193 study explored this relationship with historical reconstruction as a central motivation \cite{Marena2026DMT}. Here, we examine whether an existing solar foundation model can be adapted to the helium input and to several distinct target products. The extension is evaluated through channel-specific reconstruction and segmentation performance.

Surya provides a 366-million-parameter backbone pretrained on full-resolution SDO observations using eight AIA channels and five HMI products in a forecasting task. Its architecture combines spectral gating with long- and short-range attention \cite{Roy2025Surya}; SuryaBench describes the associated data resource and benchmark tasks \cite{Roy2025SuryaBench}. Adapting this representation requires two changes: accepting a single ground-based input and decoding the quantities relevant to the new task. We address them with a convolutional input adapter and dedicated decoders, while low-rank adaptation (LoRA) \cite{Hu2022LoRA} updates selected backbone projections. A separate residual refiner then investigates whether local corrections can improve outputs that already reproduce the broad disk structure.

The experimental sequence follows the intended historical use. Contemporary SOLIS--AIA pairs provide supervision and a downstream held-out benchmark. Validation selects among output-specific configurations, after which the retained models are applied to earlier helium observations. Observations from the Solar and Heliospheric Observatory/Extreme-ultraviolet Imaging Telescope (SOHO/EIT) provide external references for selected dates. This sequence separates learning the mapping from applying it historically: at inference, both the translator and refiner require only the helium observation and quantities generated from it.

The main contributions of this work are:
\begin{enumerate}
\item a full-disk dataset that pairs SOLIS \HeI{}, AIA 94/193/304~\AA{}, and SPoCA CH records in time and registers them to a common World Coordinate System (WCS) grid;
\item an architecture that maps single-channel \HeI{} through a convolutional adapter, independently adapted Surya backbones, and dedicated one-channel decoders;
\item validation-based hyperparameter tuning of line-specific LoRA rank, decoder architecture, and loss for AIA 94, 193, and 304~\AA{} intensity synthesis and CH probability prediction; and
\item a historical inference workflow with channel-specific residual refinement, illustrated on two pre-SDO dates and quantitatively assessed on three historical dates against independent EIT references.
\end{enumerate}

\section{Dataset and Physical Alignment}\label{sec:methods}

\subsection{Observations and scientific targets}\label{sec:data}

The input is a full-disk SOLIS Vector Spectromagnetograph He\,{\sc i} 1083.0~nm scalar product \cite{Keller2003SOLIS,NSOSOLISDataProducts}. The retained source filenames identify the observations; the precise Flexible Image Transport System (FITS) plane and physical units remain subject to the provenance limitation discussed in Section~\ref{sec:discussion}. Images are reprojected to a single-channel $4096\times4096$ solar-north-up array matching the Surya and target grid. This specifies the sampling grid, without increasing the optical resolution of the SOLIS measurement. Regression targets are level-1 AIA 94, 193, and 304~\AA{} scalar arrays obtained through the Joint Science Operations Center. AIA records $4096\times4096$-pixel EUV images at nominal 12-s cadence and approximately 0.6~arcsec per pixel \cite{Lemen2012,Boerner2012}. Each retained SOLIS record is associated with the closest admissible target in each channel.

The AIA channels are modeled as separate continuous scalar targets, not color-rendered images. CH supervision is associated with SPoCA events accessed through the Heliophysics Feature Catalogue. SPoCA partitions EUV pixels with fuzzy/possibilistic clustering and applies region extraction, tracking, and catalog characterization \cite{Verbeeck2014SPoCA}. The current manifests identify the source as \texttt{spoca\_hek} and record a maximum association window of 120~min. Accepted CH polygons are rasterized and transformed onto the same helioprojective grid as the AIA images; bilinear reprojection yields soft boundary targets in $[0,1]$. These are operational algorithmic labels, not exhaustive expert annotations. A catalog-level audit of event identifiers, constituent observations, and the distinction between verified empty and missing masks is still needed, particularly for early years.

\begin{figure*}[!htbp]
\centering
\includegraphics[width=\textwidth,height=0.32\textheight,keepaspectratio]{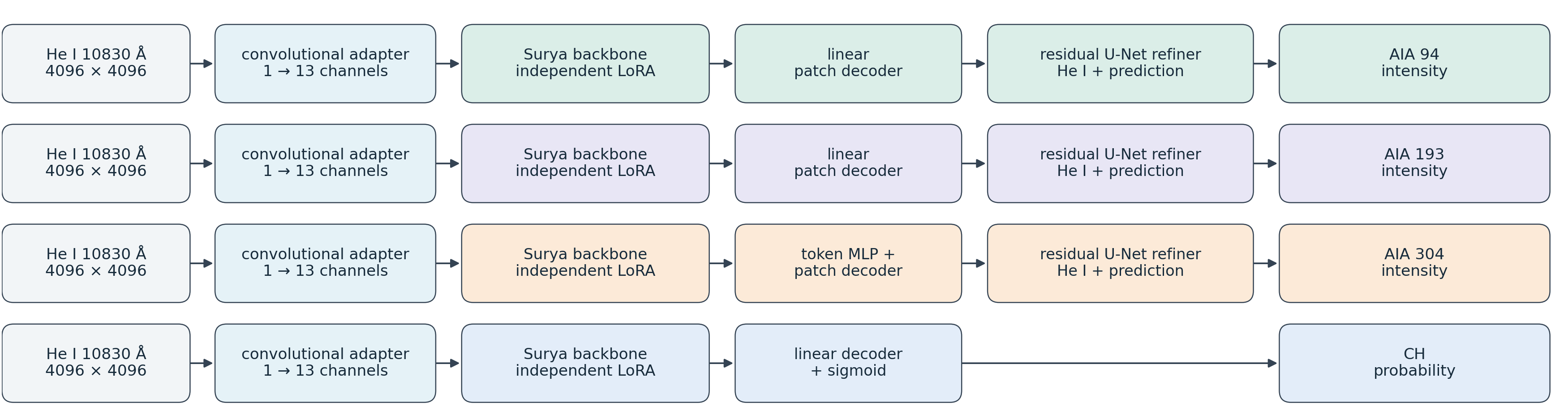}\vspace{-0.1in}
\caption{Four independent Surya--LoRA adaptations of the same full-disk He I input. Each row represents a separately tuned backbone adapter and single-output decoder. The selected AIA 304 configuration adds a nonlinear token MLP before linear patch reconstruction; the other dedicated configurations use a linear patch decoder. Each AIA branch optionally applies an independently trained residual refiner conditioned on He I and its synthetic output. The CH path converts logits to probabilities with sigmoid and bypasses refinement.}
\label{fig:architecture}\vspace{-0.2in}
\end{figure*}
\subsection{Temporal association and data partitioning}\label{sec:pairing}

Each SOLIS observation is associated with the nearest admissible AIA image in each target channel. The current matching pipeline permits an absolute SOLIS--AIA offset of at most 15~min and records the source and target timestamps, signed offset, file path, and quality status for each channel. SPoCA events are associated separately under the 120-min maximum tolerance recorded in the manifest. The pipeline assigns each sample a temporal-association weight
\begin{equation}
w_{t}=\max\!\left[0.2,\exp\!\left(-\frac{|\Delta t|}{10~\mathrm{min}}\right)\right],
\end{equation}
where $|\Delta t|$ is the largest relevant SOLIS--AIA offset for that sample. The manifest, rather than the contents of a directory, defines each experiment and makes the reported cohort auditable.

\begin{table}[!htbp]\vspace{-0.2in}
\centering
\caption{Full-resolution EUV evaluation cohorts. Counts refer to the quality-controlled manifests used by the current experiments.}\vspace{-0.1in}
\label{tab:dataset}
\scalebox{1.00}{%
\begin{tabular}{llrrr}
\toprule
Split & Months & 193~\AA{} & 304~\AA{} & 94~\AA{} \\
\midrule
Training & January--October & 679 & 679 & 678 \\
Validation & November & 87 & 87 & 87 \\
Test & December & 57 & 57 & 57 \\
\bottomrule
\end{tabular}%
}\vspace{-0.1in}
\end{table}
Random image-level splitting can place morphologically similar observations in different partitions. We instead assign January--October to training, November to validation, and December to the downstream test. The common-grid AIA 193/304 cohort contains 679 training observations from 2010--2015, 87 validation observations from 2011--2014, and 57 quality-controlled December observations on 54 distinct dates from 2012--2014; AIA 94 has one fewer training observation (Table~\ref{tab:dataset}). No explicit buffer separates November from December or December from the next January. Selection uses training and validation data only. These test observations are excluded from downstream fine-tuning and validation-based hyperparameter selection.

\subsection{World coordinate system registration and preprocessing}\label{sec:wcs}

Images are co-registered from their solar WCS metadata rather than by resizing array indices. Each modality is reprojected onto a centered, solar-north-up helioprojective grid following the coordinate conventions of \cite{Thompson2006}, using the Astropy/SunPy solar-coordinate software ecosystem \cite{Astropy2013,SunPy2020}. Native-resolution experiments retain $4096\times4096$ pixels. Scalar images are reprojected by bilinear interpolation. The catalog mask undergoes the same geometric transform and is clipped to $[0,1]$, so fractional values at reprojected boundaries are retained as soft targets. A circular evaluation support $\Omega$, centered in an $N\times N$ frame with radius $0.48N$, is stored with every sample.

Each scalar image is normalized independently using its first and 99th intensity percentiles, $q_1$ and $q_{99}$, as the lower and upper reference intensities:
\begin{equation}
I_{\mathrm{norm}}=\operatorname{clip}\left(\frac{I-q_1}{q_{99}-q_1},0,1\right).
\end{equation} The transformation maps the original intensity $I$ to $[0,1]$, clipping intensities below $q_1$ to zero and above $q_{99}$ to one.
Consequently, image metrics characterize normalized morphology and contrast, not absolute radiometric calibration; clipping also removes information in the brightest tail. This robust, scalar treatment is deliberately distinct from multiscale display normalizations developed for solar-image visualization \cite{MorganDruckmuller2014}; AIA color tables are used only for visualization. During training, rotations by multiples of $90^\circ$ and horizontal or vertical reflections are applied identically to every modality in a sample.

\section{Surya Adaptation}\label{sec:model}

Surya is pretrained on 13 SDO channels: eight AIA passbands (94, 131, 171, 193, 211, 304, 335, and 1600~\AA{}) and five HMI products. We adapt it to one full-disk \HeI{} image with a convolutional input adapter, LoRA updates to selected backbone projections, and a dedicated one-channel patch decoder. Fig.~\ref{fig:architecture} shows four separately trained models, one each for AIA 94, AIA 193, AIA 304, and CH probability. Each starts from the same pretrained Surya checkpoint but learns its own input adapter, LoRA parameters, and single-output decoder. Generating all four products therefore requires applying four independent models to the same He I observation. In a separate exploratory configuration, a single jointly trained model predicts all four targets. 

Because the present input is one \HeI{} image rather than a multichannel SDO sequence, the learned front end applies two $3\times3$ blocks combining convolution, batch normalization, and Gaussian error linear unit (GELU) activation \cite{Hendrycks2016GELU} with 32 hidden channels and a final $1\times1$ projection to 13 channels. For $x_{\rm HeI}\in\mathbb{R}^{B\times1\times H\times W}$,
\begin{equation}
x_{\rm adapt}=g_{\psi}(x_{\rm HeI})\in\mathbb{R}^{B\times13\times H\times W}.
\end{equation}
The pretrained backbone expects two time slots, so the same adapted He I image is repeated across both while preserving Surya's temporal-embedding interface. This is an architectural compatibility device; it does not create additional temporal information.

The adapted representation is encoded by Surya's HelioSpectFormer backbone. Selected linear projections are fine-tuned with LoRA while their pretrained base matrices remain frozen. For a backbone linear map $W$, a LoRA layer evaluates
\begin{equation}
Wx + \frac{\alpha}{r}BAx,
\end{equation}
where $A\in\mathbb{R}^{r\times d_{\rm in}}$ and $B\in\mathbb{R}^{d_{\rm out}\times r}$ are trainable low-rank factors and $\alpha$ controls their scale \cite{Hu2022LoRA}. We evaluate rank $r\in\{8,16,32,64\}$ for AIA 193, $r\in\{16,32,64,128\}$ for AIA 94 and AIA 304, and $r\in\{16,32,64\}$ for CH. In every rank sweep, $\alpha=2r$ and dropout is fixed at 0.05. Rank, decoder architecture, and reconstruction objective are tuned using validation performance; $\alpha$ and dropout are not tuned independently. LoRA is applied to the attention, dynamic-projection, and multilayer perceptron (MLP) projections \texttt{qkv}, \texttt{to\_dynamic\_projection}, \texttt{fc1}, \texttt{fc2}, and \texttt{proj}. The input adapter and output heads are trained jointly with these low-rank parameters.

All dedicated models use a trainable one-channel output decoder after the Surya backbone. In the linear configuration, the decoder projects each output token directly to a reconstructed image patch; it is not a nonlinear MLP. The optional \texttt{token\_mlp} configuration applies layer normalization and a residual two-layer GELU MLP in token space before the same linear patch decoder. AIA 304 uses this optional module in the selected configuration; AIA 94, AIA 193, and CH use the linear decoder. The CH decoder emits logits and applies sigmoid for probability output. LoRA, the input adapter, and the chosen decoder are trained together. The comparison concerns decoder architecture, not a trainable-versus-frozen MLP experiment.

\section{Experimental Design}\label{sec:training-evaluation}

\subsection{Training objective and model selection}\label{sec:loss}

For a target channel $c \in \mathcal{C}=\{94,193,304\}$, let $I_c(p)$ denote the normalized observed intensity at pixel $p$ and $\hat I_c(p)$ the corresponding prediction. The set $\Omega$ contains the pixels inside the solar-disk mask defined in Section~\ref{sec:wcs}, and $|\Omega|$ is the number of pixels. Only these pixels contribute to the reconstruction loss.

The base mean absolute-error (L1) loss is
\begin{equation}
\mathcal L_{\mathrm{L1},c}=\frac{1}{|\Omega|}\sum_{p\in\Omega}|\hat I_c(p)-I_c(p)|.
\end{equation}

The gradient term compares horizontal and vertical changes between adjacent pixels:
\begin{equation}
\mathcal L_{\nabla,c}=\frac12\sum_{d\in{x,y}}\frac{1}{|\Omega_d|}\sum_{p\in\Omega_d}|D_d\hat I_c(p)-D_d I_c(p)|.
\end{equation}
Here, $D_x I(i,j)=I(i,j+1)-I(i,j)$ and $D_y I(i,j)=I(i+1,j)-I(i,j)$. The set $\Omega_d$ includes only pairs whose two pixels lie inside the disk. This term encourages agreement in spatial intensity transitions rather than simply increasing image sharpness.

Optional gradient and structural terms give the channel-specific objective
\begin{equation}
\mathcal L_{\mathrm{EUV},c}=\mathcal L_{\mathrm{L1},c}+\beta_c\mathcal L_{\nabla,c}+\gamma_c\mathcal L_{\mathrm{SSIM},c}.
\end{equation}
The selected AIA 94, 193, and 304 configurations use $(\beta_c,\gamma_c)=(0.1,0)$, $(0.1,0.1)$, and $(0,0)$, respectively. The structural term penalizes local contrast and structural disagreement using structural similarity (SSIM) \cite{Wang2004SSIM}.

Per-sample objectives are aggregated as a normalized weighted mean using $w_t$ from Eq.~(1). With batch size one, this normalization cancels the sample weight; it does not reduce the contribution of less-synchronous pairs in the reported single-image batches.

For CH prediction, binary cross-entropy (BCE) on logits is combined with soft Dice loss \cite{Milletari2016VNet}. BCE penalizes pixelwise classification errors, while soft Dice penalizes disagreement in region overlap. The coefficients $w_{\rm BCE}$ and $w_{\rm Dice}$ set their relative contributions:
\begin{align}
\mathcal{L}_{\rm CH} &= w_{\rm BCE}\mathcal{L}_{\rm BCE}+w_{\rm Dice}\mathcal{L}_{\rm Dice},\\
\mathcal{L}_{\rm total} &= \sum_{c\in\mathcal{C}}\lambda_c\mathcal{L}_{\mathrm{EUV},c}
+\lambda_{\rm CH}\mathcal{L}_{\rm CH}.
\end{align}

 $\mathcal{L}_{\rm total}$ is the loss setup for CH reconstruction with $\lambda_{\rm CH}=5$, $w_{\rm BCE}=2$, and $w_{\rm Dice}=1$. Table~\ref{tab:training} summarizes the reference architecture and optimization settings. Optimization uses Adam with decoupled weight decay (AdamW) \cite{LoshchilovHutter2019}; the rank study uses a learning rate of $5\times10^{-5}$, at most 80 epochs, batch size one, mixed precision, and activation checkpointing. Checkpoint saving during training monitors the configured validation loss; after training, hyperparameter selection compares retained configurations by validation mean absolute error (MAE) for AIA outputs and by validation Dice for CH. The paper therefore reports the checkpoint actually retained by each run and the metric used to compare runs separately.

\begin{table}[!htb]\vspace{-0.2in}
\centering
\caption{Fine-tuning configuration.}\vspace{-0.1in}
\label{tab:training}
\scalebox{1.00}{%
\begin{tabular}{@{}p{0.36\columnwidth}p{0.57\columnwidth}@{}}
\toprule
Component & Setting \\
\midrule
Input/output grid & $4096\times4096$ pixels \\
Input adapter & $1\rightarrow32\rightarrow32\rightarrow13$ channels \\
Output decoder & linear patch projection; optional token MLP \\
LoRA targets & attention, dynamic, and MLP projections \\
LoRA rank/scale & varied by output; $\alpha=2r$ in sweeps \\
LoRA dropout & 0.05 \\
Optimizer & AdamW \\
Rank-study learning rate & $5\times10^{-5}$ \\
Maximum epochs & 80\\
Early-stopping patience & 12 consecutive epochs without sufficient validation improvement\\
Batch size & 1 \\
Numerical strategy & mixed precision, checkpointing \\
\bottomrule
\end{tabular}%
}\vspace{-0.10in}
\end{table}

\begin{table*}[!htb]
\centering
\caption{Native-resolution hyperparameter tuning for dedicated outputs. Each row keeps its stated head and objective fixed. AIA entries are validation MAE (lower is better); CH entries are validation Dice (higher is better). L1 denotes absolute-error loss, G gradient consistency, and BCE binary cross-entropy. Dashes denote rank--configuration combinations without a reported validation score; bold entries identify the best reported score in each row.}\vspace{-0.1in}
\label{tab:rank-validation}
\scalebox{1.00}{%
\begin{tabular}{lllrrrrr}
\toprule
Output & Fixed configuration & Metric & $r=8$ & $r=16$ & $r=32$ & $r=64$ & $r=128$\\
\midrule
AIA 94 & Linear, L1 & MAE $\downarrow$ & -- & 0.0648 & 0.0660 & \textbf{0.0637} & --\\
AIA 94 & Linear, L1+G & MAE $\downarrow$ & -- & -- & -- & \textbf{0.0628} & 0.0723\\
AIA 193 & Linear, L1 & MAE $\downarrow$ & 0.0578 & 0.0566 & \textbf{0.0546} & 0.0558 & --\\
AIA 304 & Linear, L1 & MAE $\downarrow$ & -- & 0.0730 & 0.0764 & \textbf{0.0724} & --\\
AIA 304 & Token MLP, L1 & MAE $\downarrow$ & -- & -- & -- & \textbf{0.0720} & 0.0788\\
CH & Linear, $2$BCE+$1$Dice & Dice $\uparrow$ & -- & \textbf{0.4162} & 0.4102 & 0.4100 & --\\
\bottomrule
\end{tabular}%
}\vspace{-0.2in}
\end{table*}
\subsection{Experimental controls and reproducibility}\label{sec:controls}

Each rank-sweep row changes the declared rank while retaining its manifest, grid, normalization, loss, and decoder. Separate loss and decoder follow-ups change other components, as marked in Table~\ref{tab:rank-validation}; their scores are not interpreted as a pure rank effect. Run logs record random seeds, arguments, and validation results. Test metrics for candidates at different rendering resolutions are not pooled. We do not use a matched random-initialization or small-network baseline to estimate how much of the achieved accuracy is due specifically to Surya pretraining.

The comparison between joint and dedicated outputs is descriptive: the two settings have different optimization objectives, so it does not isolate a causal effect of task count. The older full-weight and frozen-backbone development runs used different data cohorts and are excluded from the present native-resolution comparison. All headline AIA scores refer to normalized scalar arrays, not red--green--blue (RGB) rendering.

\subsection{Active-region-aware residual refinement}\label{sec:refinement}

The dedicated Surya--LoRA models recover the broad organization of the EUV disk but can suppress compact bright cores and fine active-region structure. A high full-disk correlation can therefore coexist with scientifically important local errors. Rather than asking the global translator to relearn these details, we use separate residual attention U-Nets for AIA 94, 193, and 304~\AA{} on scalar arrays, adapting the U-Net encoder--decoder and gated skip-connection ideas \cite{Ronneberger2015UNet,Oktay2018AttentionUNet}. Patch-based training concentrates capacity on local structure; active-region-focused sampling and loss weighting emphasize the features that the first stage most often smooths. Each model receives the normalized He I input $I_{\rm HeI}$ and frozen Surya prediction $\hat I_{\rm Surya}$ and predicts a correction $R_{\phi}$,
\begin{equation}
\hat{I}_{\rm ref}=\operatorname{clip}\!\left[
\hat{I}_{\rm Surya}+R_{\phi}\!\left(I_{\rm HeI},\hat{I}_{\rm Surya}\right),0,1\right].
\end{equation}
The residual formulation separates global prediction from local correction: the network learns an additive adjustment rather than synthesizing the disk from scratch. Conditioning on He I retains access to the observation instead of relying exclusively on structures already present in the synthetic image. Neither choice guarantees that newly sharpened features are physically correct. Refinement uses overlapping $512\times512$ tiles and blended reconstruction. Pixels outside the WCS solar-disk support $\Omega$ are excluded.

The refiner is trained only on paired SDO-era training observations. HMI Active Region Patch (HARP) masks identify active-region tiles for oversampling; they are not an inference input \cite{Bobra2014SHARPs}. The four-level network uses a base width of 48 channels, $512\times512$ tiles, and skip-connection attention gates. The training sampler targets active-region tiles with probability 0.5 and hard examples with probability 0.25. Its disk-weighted robust pixel loss uses weight $1+4M_{\rm AR}+2M_{\rm bright}$, where the bright mask is defined above the 95th percentile of the target tile; a gradient-consistency loss has coefficient 0.1. AdamW starts at $2\times10^{-4}$. Training runs for at most 80 epochs and stops early after 12 consecutive epochs without a validation-loss improvement meeting the configured minimum threshold. Overlapping tiles are blended during full-disk inference. No EIT image or historical target enters refiner training; historical refinement consumes only He I and a synthetic AIA prediction, allowing the same inference pipeline to operate where no AIA observation or HARP mask is available.

\begin{table*}[!htbp]
\centering\setlength{\tabcolsep}{4pt}
\caption{Native-resolution held-out performance without refinement. AIA entries are mean image-level scores from $4096\times4096$ scalar outputs on 57 December observations and only within the WCS solar disk. Each row states the exact retained candidate configuration. G and S denote gradient and SSIM losses, each with weight 0.1.}
\label{tab:results}\vspace{-0.1in}
\scalebox{1.0}{%
\begin{tabular}{lrrlrrrrrr}
\toprule
Output & Rank & $\alpha$ & Head/objective & CC & MAE & PSNR & SSIM & F1 & IoU\\
\midrule
AIA 94 & 64 & 128 & Linear, L1+G & 0.8196 & 0.0597 & 21.193 & 0.2806 & -- & --\\
AIA 193 & 32 & 64 & Linear, L1+G+S & 0.8885 & 0.0443 & 22.716 & 0.8532 & -- & --\\
AIA 304 & 64 & 128 & Token MLP, L1 & 0.8398 & 0.0606 & 20.775 & 0.5317 & -- & --\\
CH & n.r. & n.r. & Linear + sigmoid & -- & -- & -- & -- & 0.6073 & 0.4360\\
\bottomrule
\end{tabular}%
}
\par\smallskip\parbox{0.97\textwidth}{ A dash denotes an inapplicable metric; n.r. means the checkpoint metadata was not recorded. CH results are from an earlier dedicated checkpoint rather than the displayed CH rank sweep.}\vspace{-0.2in}
\end{table*}
\subsection{Evaluation protocol}\label{sec:evaluation}

Continuous outputs are evaluated on normalized scalar arrays using Pearson correlation coefficient (CC), mean absolute error (MAE), peak signal-to-noise ratio (PSNR; unit data range), and structural similarity (SSIM) \cite{Wang2004SSIM}. A fixed WCS disk of radius $0.48N$ excludes off-disk pixels; SSIM averages only locations whose complete $7\times7$ local window lies inside this disk. The evaluation summaries average image-level statistics across the 57 observations. CH Dice/F1 and intersection over union (IoU) threshold predictions and catalog targets at 0.5 and pool on-disk confusion counts. Calendar-day-cluster bootstrap confidence intervals are reported for the AIA test metrics; date-level CH intervals are not available. These intervals do not treat millions of correlated pixels as independent samples. Any metrics on color-mapped RGB renderings belong to a separate display-dependent protocol.

\section{Results}\label{sec:results}

We first compare configurations using November validation MAE for the AIA outputs and validation Dice for CH. We then evaluate the retained checkpoints on the 57-observation December cohort at a $4096\times4096$ output grid. Hyperparameter tuning and test evaluation use different splits and serve different purposes; checkpoint availability is not treated as a selection metric.

\begin{figure*}[!htbp]
\centering
\includegraphics[width=\textwidth,height=0.90\textheight,keepaspectratio]{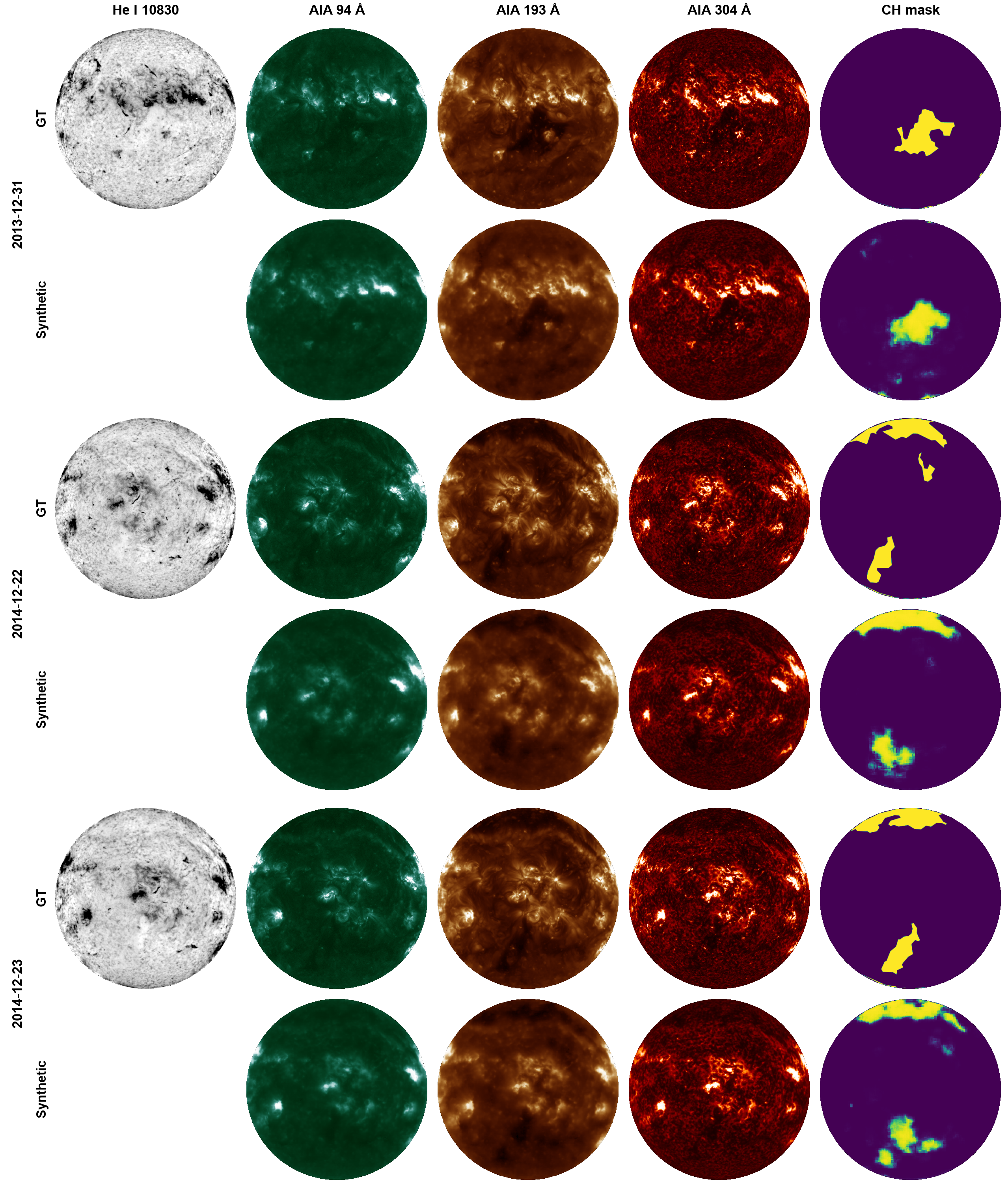}
\caption{Archived illustrative held-out examples from earlier dedicated Surya--LoRA checkpoints; these are not the exact AIA 94/304 rank-64 checkpoints in Table~\ref{tab:results}. Columns: observed \HeI{}, AIA 94, AIA 193, AIA 304, and CH. For each date, the upper row contains observations/catalog targets, and the lower row contains synthetic outputs; the synthetic \HeI{} cell is blank. The underlying predictions are native $4096\times4096$ scalar arrays; color mapping and print rasterization are used only for display. }
\label{fig:examples}
\end{figure*}

\subsection{Output-specific hyperparameter tuning}\label{sec:results-tuning}

We examine LoRA rank separately for each dedicated output. The search covers selected combinations of rank, decoder, and loss rather than all possible combinations. Table~\ref{tab:rank-validation} summarizes native-resolution validation scores from the hyperparameter search. Within each row, the stated decoder and objective are fixed. AIA channels are selected by validation MAE and CH by validation Dice. No independent effect is attributed to $\alpha$ or dropout because $\alpha=2r$ and dropout is fixed at 0.05 in the rank sweeps.

Under the fixed linear-L1 comparison, the lowest validation MAE occurs at rank 64 for AIA 94/304 and rank 32 for AIA 193. In separate follow-ups, gradient regularization further improves AIA 94 at rank 64, the tuned AIA 193 checkpoint uses gradient plus SSIM regularization at rank 32, and a token MLP improves AIA 304 at rank 64. The displayed CH rank sweep favors rank 16, but its loss weights and provenance differ from the older dedicated CH checkpoint reported in Table~\ref{tab:results}. Raising the tuned AIA 94/304 rank to 128 worsens validation MAE in the reported single runs; this alone does not establish overfitting.

\subsection{Selected checkpoints and held-out performance}\label{sec:results-primary}

Table~\ref{tab:results} reports the retained dedicated checkpoints on the common native-grid test set. The selected AIA 193 model provides the strongest structural match, whereas the low AIA 94 SSIM indicates remaining local structural discrepancies, without identifying their cause. The AIA scores are means of 57 image-level disk metrics. For AIA 94, 193, and 304, respectively, the 95\% calendar-day-cluster bootstrap intervals for CC are [0.8061, 0.8324], [0.8836, 0.8931], and [0.8298, 0.8500], and for SSIM are [0.2610, 0.3011], [0.8482, 0.8581], and [0.5183, 0.5458]. We sampled the 54 distinct test dates with replacement 10,000 times, preserving repeated observations on a date as a cluster. These intervals capture test-date sampling uncertainty, not systematic label error or training-seed variability. The dedicated CH score comes from a separate earlier retained checkpoint; its complete rank/scale metadata and date-level counts are unavailable, so no confidence interval or tuned-rank attribution is given.

\subsection{Spatial fidelity and the motivation for refinement}\label{sec:results-qualitative}

The metrics in Table~\ref{tab:results} distinguish several aspects of reconstruction fidelity. CC measures whether spatial variations track the target; MAE and PSNR quantify normalized intensity discrepancies; SSIM additionally responds to local contrast and structure. Their joint interpretation matters because a smooth prediction can reproduce the locations of bright regions without resolving their internal morphology. AIA 193 performs strongly under both global correlation and local structural measures. The larger separation between CC and SSIM for AIA 94, and to a lesser extent AIA 304, identifies remaining structural error without establishing its physical or architectural cause. For CH, overlap metrics instead assess the area and boundary agreement of thresholded regions, not the reconstruction of EUV texture.

Fig.~\ref{fig:examples} compares the He I inputs, observed AIA targets, SPoCA-derived CH targets, and dedicated-model predictions on 2013 December 31, 2014 December 22, and 2014 December 23. Each date occupies two rows: observations and targets above, synthetic outputs below; the lower He I cell is intentionally blank. Across these three illustrated dates, prominent active regions and broad dark structures appear in approximately the observed locations. The synthetic EUV maps nevertheless smooth fine emission patterns, especially within active regions, and the predicted CH boundaries remain less precise than the catalog targets. This contrast between recovered global organization and unresolved local detail motivates a residual refiner that starts from, rather than replaces, the Surya prediction. The figure dates were selected for illustration, not sampled randomly. Moreover, the archived figure was generated from earlier dedicated checkpoints; its AIA 94 and 304 panels are not predictions from the exact rank-64 checkpoints in Table~\ref{tab:results}. It therefore illustrates model-family behavior rather than visually validating differences among the selected checkpoints.

\subsection{Historical synthesis and cross-instrument refinement}\label{sec:eit-audit}

The paired SOLIS--AIA cohort remains the primary quantitative benchmark. To examine the historical use case, we also apply the retained AIA 193 and 304 Surya--LoRA models and their channel-specific residual refiners to earlier SOLIS observations with SOHO/EIT 195 and 304~\AA{} references. EIT 195 is not the same passband as AIA 193, and the 304~\AA{} instruments also have different responses \cite{Delaboudiniere1995EIT,Marena2026DMT}. EIT is therefore an independent cross-instrument reference, not AIA ground truth.

\begin{figure*}[!htbp]
\centering
\includegraphics[width=\textwidth,height=0.9\textheight,keepaspectratio]{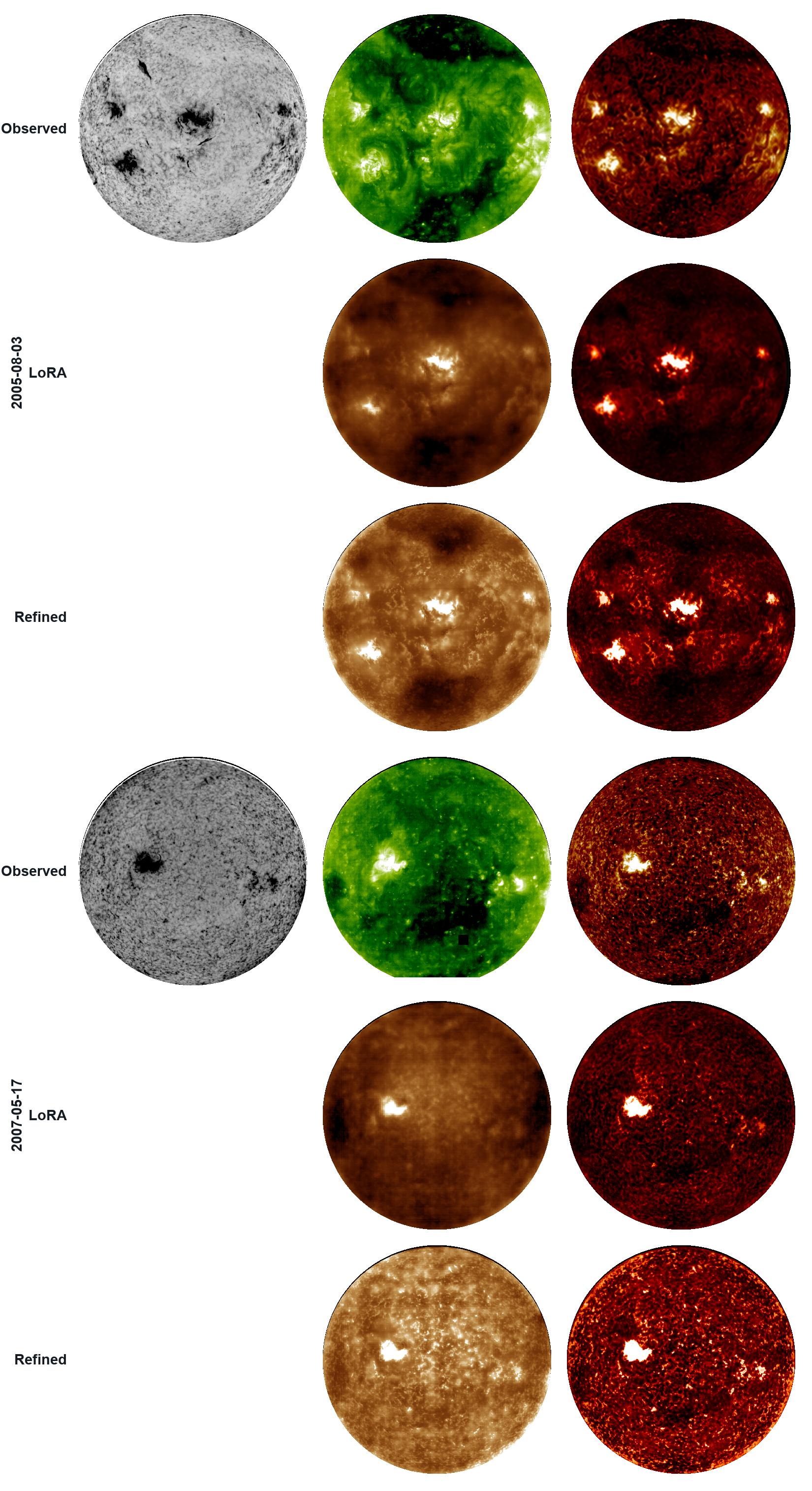}
\caption{Historical synthesis for 2005 August 3 and 2007 May 17. Each date occupies three rows: observations, Surya--LoRA predictions, and refined predictions. The columns contain He I 10830~\AA{}, EIT 195/synthetic AIA 193~\AA{}, and EIT 304/synthetic AIA 304~\AA{}. The layout contains six rows and three columns. He I appears only in the first row of each date group; its two lower cells are blank. Predictions are generated as $4096\times4096$ scalar arrays; colors are for display. EIT provides an independent cross-instrument reference, not AIA ground truth.}
\label{fig:historical-eit}
\end{figure*}

We identify 27 records before 2010 May 1 for which both EIT channels are within 15~min of the He I observation. Historical SOLIS images are reprojected to the same centered, solar-north-up $4096\times4096$ grid and solar-disk radius used during training, and are normalized with the unchanged scalar first/99th-percentile procedure. The EIT Level-1 products are processed according to their available calibration metadata, without treating their displayed color scales as physical units; all EIT and AIA colour mappings are used for display only. For each date, Surya first produces scalar AIA 193 and AIA 304 arrays. The corresponding channel-specific refiner then predicts a residual correction using the He I input and the frozen Surya output. Thus, the refined historical images are obtained without EIT fitting, histogram matching, or EIT-derived supervision.

\begin{table}[!htb]\vspace{-0.2in}
\centering
\caption{Exploratory agreement of synthetic AIA 304~\AA{} with observed SOHO/EIT 304~\AA{} on three historical dates (2005 July 10, 2005 August 3, and 2007 May 17); the latter two are illustrated in Fig.~\ref{fig:historical-eit}. Mean image-level metrics use a common $1024\times1024$ grid and only the solar disk. EIT is a cross-instrument reference, not AIA ground truth.}\vspace{-0.1in}
\label{tab:historical-eit304}
\scalebox{1.00}{%
\begin{tabular}{lrrrr}
\toprule
Method & CC $\uparrow$ & MAE $\downarrow$ & PSNR $\uparrow$ & SSIM $\uparrow$ \\
\midrule
Surya--LoRA & 0.6511 & 0.1032 & 16.985 & 0.3610 \\
Surya--LoRA + refiner & \textbf{0.7085} & \textbf{0.0668} & \textbf{20.597} & \textbf{0.4535} \\
\bottomrule
\end{tabular}%
}\vspace{-0.1in}
\end{table}
Fig.~\ref{fig:historical-eit} compares observations, direct synthesis, and refinement for 2005 August 3 and 2007 May 17. Each date occupies three rows: the He I input and EIT observations, the direct Surya--LoRA outputs, and the refined outputs. The second column compares EIT 195 with synthetic AIA 193, while the third compares EIT 304 with synthetic AIA 304. The EIT panels provide cross-instrument references rather than identical-passband AIA targets. The AIA 304--EIT 304 comparison in Table~\ref{tab:historical-eit304} provides an exploratory quantitative complement to the visual examples. Refinement increases CC from 0.6511 to 0.7085 and SSIM from 0.3610 to 0.4535, decreases MAE from 0.1032 to 0.0668, and increases PSNR from 16.985 to 20.597~dB. The reported means retain the original three-date evaluation cohort (2005 July 10, 2005 August 3, and 2007 May 17), of which two dates are displayed, and use a $1024\times1024$ common-grid solar-disk evaluation; they are not pooled with the native-grid AIA test results in Table~\ref{tab:results}. No corresponding EIT 195 metric or historical CH score is claimed.

\section{Discussion}\label{sec:discussion}

The principal result is that a single full-disk \HeI{} observation contains information from which dedicated models can predict substantial portions of the morphology visible in multiple EUV passbands and in CH maps. The outputs are not interchangeable: AIA 193 shows the strongest held-out structural agreement, whereas the lower AIA 94 SSIM reveals that global correlation does not guarantee recovery of fine hot-coronal texture. These differences support treating the wavelengths as distinct synthesis tasks rather than assuming one generic EUV representation suffices.

The historical significance is that the trained model requires only a He I observation at inference. It can therefore generate synthetic AIA-like morphological proxies for dates preceding direct AIA imaging, extending the question explored in our earlier diffusion-based He I-to-AIA 193 study \cite{Marena2026DMT} to separate 94, 193, and 304~\AA{} outputs and a CH product. The goal is a common representation with which to examine broad solar structure across observational eras, not to replace archival observations or uniquely invert the atmospheric state. In the two illustrated pre-SDO examples, co-temporal EIT images provide independent context for the locations of bright and dark structures.

The two-stage pipeline addresses different spatial scales. Surya--LoRA recovers the full-disk arrangement, while the patch-based residual refiner is intended to restore local contrast and texture that the first stage smooths. In the three-date AIA 304--EIT 304 comparison, refinement improves all four reported metrics, including CC from 0.6511 to 0.7085 and SSIM from 0.3610 to 0.4535. This is evidence of improved agreement with an independent historical reference on those dates, not evidence that every added fine structure is physically correct. WCS registration and disk-restricted scoring remain essential: they prevent pointing differences and extensive off-disk background from masquerading as model performance. Likewise, the comparison of joint and dedicated objectives is descriptive rather than a causal demonstration that task separation alone produced the gains.

The hyperparameter search also informs how this workflow should be extended. Increasing adaptation capacity is not uniformly beneficial: the reported rank-128 follow-ups do not improve on their rank-64 counterparts, while AIA 193 favors rank 32 in its linear-head sweep. This supports validation-based selection for each output rather than using the largest rank throughout. The refiner is a complementary intervention, not another rank setting: it changes the spatial scale and emphasis of the learning problem while retaining the translator as a fixed first stage. Consequently, its added value must be judged against the corresponding unrefined predictions on the same observations, with identical registration, normalization, and metric support.

Several limitations constrain the interpretation. The primary test contains only 54 unique December dates and has no temporal buffer at month boundaries. The SOLIS input plane and physical units require FITS-header verification; per-image normalization discards absolute radiometric information. SPoCA targets are algorithmic labels whose early-year provenance and missing-versus-empty status need further review. We lack a matched random-initialization baseline, date-level CH confidence intervals, and a checkpoint-matched full-test evaluation of the refiner using the same metric implementation as Table~\ref{tab:results}. The historical EIT comparison spans only three selected dates at a different evaluation grid and includes instrument-response, resolution, and quality-flag differences; it cannot validate AIA-equivalent fine detail or broad historical generalization. These audits and controlled evaluations are prerequisites for stronger physical claims.

\section{Conclusions}\label{sec:conclusions}

We presented a cross-modal framework that adapts Surya to synthesize full-disk AIA 94, 193, and 304~\AA{} intensity proxies and CH probability maps from a single \HeI{} observation. A convolutional input adapter connects the ground-based observation to the pretrained representation, while LoRA updates and dedicated decoders specialize that representation for each output. Validation-based hyperparameter tuning identifies different preferred configurations across channels, demonstrating the practical value of output-specific model selection rather than adopting one rank and objective for every target.

On the downstream held-out December cohort, the selected unrefined models achieved disk-restricted CCs of 0.8196, 0.8885, and 0.8398 for AIA 94, 193, and 304, respectively. The separately retained CH model reached F1 $=0.6073$ and IoU $=0.4360$. These results establish that the same helium input can support multiple complementary EUV reconstructions and an algorithmically supervised segmentation product. The distinction between strong global correlation and channel-dependent structural agreement also motivates the local refinement stage rather than treating full-disk correlation alone as sufficient evidence of image fidelity.

A central outcome is the ability to apply the trained workflow to historical helium observations without requiring a contemporary AIA image. The translator and channel-specific residual refiner together provide a practical route from an archival ground-based observation to a common space-based imaging representation. On the three-date historical evaluation, refinement improved synthetic AIA 304 agreement with independent EIT 304 references from CC $=0.6511$ to $0.7085$ and SSIM $=0.3610$ to $0.4535$. The two illustrated dates show how direct synthesis and refinement can be compared with historical observations within the same workflow, without using those references as inference inputs.

The framework provides a foundation for extending multichannel solar-morphology studies to periods preceding AIA, complementing the historical measurements. Its present products are normalized morphological proxies, not calibrated radiance measurements or uniquely determined atmospheric states. Extending the historical evaluation, assessing active-region fidelity on matched full-test predictions, and quantifying uncertainty are the next steps toward a more comprehensive reconstruction of the solar record.

\section*{Acknowledgment}
SOLIS data used in this work are produced cooperatively by the National Science Foundation/National Solar Observatory (NSF/NSO) and the National Aeronautics and Space Administration/Living With a Star (NASA/LWS) program. SDO data are courtesy of NASA/SDO and the AIA science team. SOHO is a project of international cooperation between the European Space Agency (ESA) and NASA. The authors acknowledge the Royal Observatory of Belgium for the SPoCA coronal-hole catalogue and the Heliophysics Integrated Observatory (HELIO) Heliophysics Feature Catalogue infrastructure used to access feature records. This research was supported by NASA grants 80NSSC24M0174 and 80NSSC26K1242. This work used high-performance computing resources provided by the New Jersey Institute of Technology.

\bibliography{references}
\bibliographystyle{IEEEtran}

\end{document}